\documentclass[graybox]{svmult}

\usepackage{type1cm}        
\usepackage{makeidx}         
\usepackage{graphicx}        
\usepackage{multicol}        
\usepackage[bottom]{footmisc}

\usepackage{newtxtext}       %
\usepackage[varvw]{newtxmath}       

\newcommand{\mnras}{Mon.\ Not.\ R.\ Astron.\ Soc.}
\newcommand{\apj}{Astrophys.\ J.}
\newcommand{\apjl}{Astrophys.\ J.\ Lett.}

\makeindex             
                       
 \def\lsim{\lower.5ex\hbox{$\; \buildrel < \over \sim \;$}}
\def\gsim{\lower.5ex\hbox{$\; \buildrel > \over \sim \;$}}
\def\simeq{\lower.3ex\hbox{$\; \buildrel \sim \over - \;$}}

\def\rs{R_{\rm s}}

\def\bob{B_{\rm ob}}
\def\thvir{\Theta_{\rm vir}}
\def\thob{\Theta_{\rm ob}}
\def\tvir{T_{\rm vir}}
\def\tob{T_{\rm ob}}
\def\rob{r_{\rm ob}}
\def\amk{\lambda_{\rm K}}
\def\amob{\lambda_{\rm ob}}

\begin{document}

\title*{Study of accretion disks around black holes with two types of gas inflows}
\author{Rajiv Kumar\orcidID{0000-0001-9928-2770}}
\institute{Rajiv Kumar \at Harish-Chandra Research Institute, Chhatnag road, Jhunsi, Prayagraj-211019, India, \email{rajivkumar@hri.res.in}
}
%
%
\maketitle
\abstract{ We utilized a steady-state, axisymmetric, viscous hydrodynamic fluid equations around a black hole in Schwarzschild geometry background. Here, the relativistic Schwarzschild geometry is mimicked by the Paczy{\'n}sky-Wiita potential. We investigated two types of inflowing gases that can generate different kinds of accretion flows around the central objects. The inflowing gases are presented on the local energies ($B_{ob}$) of the gases versus the outermost accretion boundary locations ($r_{ob}$) plane we named it the outermost boundary condition (OBC)-plane. 
Based on the energies of the inflowing gases we found two types of inflowing gas classified as cold-mode and hot-mode inflowing gases in the OBC-plane. Doing so we have found the initial temperature of the inflowing gases can be a parameter for the study of the accretion process. 
As it can affect the disk structure and optical depth of the accretion flow, which in turn can impact the radiative emissions observed in many accreting sources.
}

\section{Introduction}
\label{sec:intro}
Accretion is a fundamental process in astrophysics that describes the accumulation of matter, typically gas or dust, onto a massive object due to gravity \cite{B52}. As material falls onto the central object, it forms a rotating disk known as an accretion disk. These disks are common in various astrophysical systems, including young stellar objects, black holes, and active galactic nuclei. Accretion disks can be extremely hot and luminous, emitting radiation across the electromagnetic spectrum. The physics governing the
evolution of accretion disks involves processes such as viscosity and angular
momentum transport \cite{ss73}. Viscosity within the disk causes the material to lose angular
momentum and spiral inward towards the central object. The exact mechanism
responsible for viscosity in accretion disks remains an area of active research, with
turbulence, magnetic fields, and other effects playing potential roles. 

The accretion rate at which material accretes onto the central object can
vary significantly depending on factors such as the mass of the central object, the
properties of the surrounding material, and the efficiency of angular momentum
transport within the accretion disk \cite{kg19,ky21}.  {Usually, the accretion rate is a free parameter, however, in extreme cases, it can be limited by the Eddington luminosity limit which depends on the mass of the central object.}
Accretion rates are crucial for understanding the
growth of stars, black holes, and other celestial bodies. In the presented study, we will study the qualitative aspects of the accreting gases which are defined at the outer-accretion boundary location (OBL), such as the initial temperature, compositions, magnetization level, ionization level, and initial angular momentum of the gases. Due to the limitations of the present model equations, we will focus only on the initial temperature of the inflowing gases around the black hole. With the changing temperature of inflowing gas, we have investigated many kinds of accretion solutions.
This article is organized in the next section \ref{sec:2} governing equations and assumptions,  classification of inflowing gases is described in section \ref{sec:3}, results are discussed in section \ref{sec:4}, and finally concludes this work in section \ref{sec:5}.

\section{Governing equations and assumptions}
\label{sec:2}
In this study, we focused only on the fluid properties of the accretion disk and corresponding OBCs.
We assumed the axis-symmetric, steady state, and viscous hydrodynamic fluid flow around a non-rotating black hole (BH).
The space-time properties around the non-rotating BH are described by the pseudo-Newtonian potential \cite{pw80}. 
We used the geometric unit system to represent the fluid equations of motion and flow quantities where $2G=M=c=1$. Here, $M$ is the mass of the BH, $c$ is the speed of light, and 
$G$ is the Gravitational constant. So, the units of length,
mass, speed and time are the Schwarzschild radius or $\rs=2GM/c^2$, $M$, $c$, and
$\rs/c=2GM/c^3$, respectively. 
\subsection{Equations of motion for accretion flow}
\label{subsec:2}
The equations of motion for viscous hydrodynamic flow on the equatorial plane
in cylindrical coordinates ($r$, $\phi$, $z$) are described by,

\noindent the radial-momentum equation:
\begin{equation}
v\frac{dv}{dr} + \frac{1}{\rho}\frac{dp}{dr} -\frac{\lambda^2}{r^{3}} + \frac{1}{2(r-1)^2}= 0
\label{rme.eq}
\end{equation}
The azimuthal-momentum equation:
\begin{equation}
v\frac{d\lambda}{dr}+\frac{1}{\Sigma r}\frac{d(r^{2}W_{r\phi})}{dr}=0
\label{amde.eq}
\end{equation}
The $z-$momentum equation with considering the disk in hydrostatic equilibrium in $z-$direction, which gives us the local disc half-height expression:
\begin{equation}
h=\sqrt{\frac{2}{\gamma}}ar^{1/2}(r-1)
\label{hh.eq}
\end{equation}
The mass-accretion rate equation:
\begin{equation}
\dot{M}=2\pi\Sigma v r,
\label{mf.eq}
\end{equation}
here, ${\dot M}$ is a constant of motion.
The entropy generation equation:
\begin{equation}
\Sigma v T\frac{ds}{dr}=f_{\rm adv}~Q^{+}.
\label{ege.eq}
\end{equation}
The local variables $v,~a,~p,~\rho$ and $\lambda$ in the above equations are the radial bulk velocity,
sound speed, isotropic pressure, gas density, and specific angular momentum (AM) of the flow, respectively.
The local viscous heating therm is given by $Q^{+}=W_{r\phi}^{2}/{\eta}$, and we assumed radiative inefficient flow, so advection factor $f_{\rm adv}=1$ in this study. 
Here, $\Sigma=2\rho h$ and $W_{r\phi}=\eta r\frac{d\Omega}{dr}$ are the vertically integrated density and the viscous stress tensor. 
 $\eta=\rho\nu h$ is the dynamic viscosity coefficient, 
$\nu=\alpha a^{2}/(\gamma \Omega_{k})$ is the kinematic viscosity, $\alpha$ is the
viscosity parameter, $\Omega$ and $\Omega_{k}$ are
the local angular velocity and local Keplerian angular velocity, respectively.
Other quantities,  $s$, $T$, and $h$ are denoted for the entropy density, the local
temperature in the physical unit, and the local half-height of the disk, respectively. 

The energy per unit mass can be obtained by integrating equation (\ref{rme.eq}) with the help of equations (\ref{amde.eq}, \ref{mf.eq}, and \ref{ege.eq}), which yields the constant of motion of the flow.
\begin{equation}
E =\frac{v^{2}}{2} + \frac{a^{2}}{\gamma-1}-\frac{\lambda^{2}}{2r^{2}}+\frac{\lambda_{0}\lambda}{r^{2}}
-\frac{0.5}{r-1} ,
\label{ge.eq}
\end{equation}
where, E is called the global specific energy of the flow, and the constant throughout the accretion flow even
in the presence of viscosity \cite{bl03,gl04}.  The local specific energy of the accretion flow can be given by the canonical Bernoulli parameter ($B$) that is a constant of motion for inviscid flow and defined as,
\begin{equation}
B=\frac{v^{2}}{2} + \frac{a^{2}}{\gamma-1}+\frac{\lambda^{2}}{2r^{2}}
-\frac{0.5}{r-1} .
\label{be.eq}
\end{equation}

The adiabatic sound speed is defined as 
\begin{equation}
a=\sqrt{\frac{\gamma p}{\rho}}=\sqrt{\frac{2\gamma\Theta}{\tilde{t}}},
\label{as.eq}
\end{equation}
where $\gamma$ is an adiabatic index, $\Theta$ is a dimensionless temperature of the gas, and $\tilde{t}=1+m_{p^+}/m_{e^-}$.

Integrating eq. (\ref{amde.eq}) with the help of eq. (\ref{mf.eq}) and the expression of angular velocity gradient is obtained,
\begin{equation}
\frac{d\Omega}{dr}=- \frac{\gamma v\Omega_{k}(\lambda-\lambda_{0})}{\alpha a^{2}r^{2}}.
\label{daf.eq}
\end{equation}
where $\lambda_{0}$ is specific AM at the BH event horizon obtained by considering
vanishing torque at the horizon \cite{bl03}. Moreover, $\Omega_{k}$ denotes the Keplerian angular velocity. 
Since $\lambda=r^{2}\Omega$, the radial derivative of $\lambda$ can be written by
\begin{equation}
\frac{d\lambda}{dr}=2r\Omega+r^{2}\frac{d\Omega}{dr}.
\label{dsam.eq}
\end{equation}

Simplified Eqs (\ref{rme.eq}, \ref{amde.eq}  and  \ref{ege.eq}), with the help of Eqs (\ref{hh.eq}, and \ref{daf.eq}) we obtain,
\begin{equation}
\frac{dv}{dr}=\frac{N}{D}.
\label{du.eq}
\end{equation}
where, 
$$
N=\frac{2}{\gamma+1}\frac{(5r-3)v}{2r(r-1)}+\frac{(\lambda^{2}-\lambda_{k}^{2})v}{a^{2}r^{3}}
+\gamma^{2}\left(\frac{\gamma-1}{\gamma+1}\right)\frac{v^{2}\lambda_{k}(\lambda-\lambda_{0})^{2}}
{\alpha a^{4}r^{4}}
$$ and
$$
D=\frac{v^{2}}{a^{2}}-\frac{2}{\gamma+1}
$$
The gradient of sound speed is obtained,
\begin{equation}
\frac{da}{dr}=\left(\frac{a}{v}-\frac{\gamma v}{a}\right)\frac{dv}{dr}+\frac{(5r-3)a}{2r(r-1)}
+\frac{\gamma (\lambda^{2}-\lambda_{k}^{2})}{ar^{3}}.
\label{da.eq}
\end{equation}
The accretion disk solutions in vertical equilibrium are solved by simultaneously integrating Eqs.
(\ref{dsam.eq}, \ref{du.eq}, \ref{da.eq}) using Runge-Kutta 4th order numerical method.  We are following the same methodology as used in these studies \cite{kc13,kg19}. 
In doing so, firstly, we have investigated a critical point (CP) of the flow corresponding to the given parameters ($E, \lambda_0, \alpha$) by the iteration mothed with the help of the CP conditions as described in those studies. The CP conditions are obtained from the Eq. \ref{du.eq} by setting $N=0$ and $D=0$.
As we know the accretion flow can have multiple critical points (MCPs), which can have shocks. So we have also investigated the adiabatic shock conditions in the accretion flows which can give us the shock location and outer CP of the accretion flow if it occurs.

The Rankine-Hugoniot shock conditions are obtained by conservation of the momentum flux, mass flux, and energy
flux across the shock. The shock conditions are given by,
the mass flux conservation,
\begin{equation}
\dot{M_{+}}=\dot{M_{-}},
\label{mfs.eq}
\end{equation}
The momentum conservation,
\begin{equation}
W_{+}+\Sigma_{+} v_{+}^{2}=W_{-}+\Sigma_{-} v_{-}^{2},
\label{rmf.eq}
\end{equation}
and the third shock condition is the energy conservation,
\begin{equation}
E_{+} =E_{-}.
\label{ef.eq}
\end{equation}
Here, subscripts minus(-) and plus(+) of the flow variables denote the quantities before and after the shock, respectively. $W$ is the vertically integrated pressure. 
We used shock conditions equations (\ref{mfs.eq}-\ref{ef.eq}) and found out the shock locations with outer CP in the global accretion solutions. 

%
%
\section{OBC-plane}
\label{sec:3}
\begin{figure}[b]
\sidecaption
\includegraphics[angle=0,scale=.66]{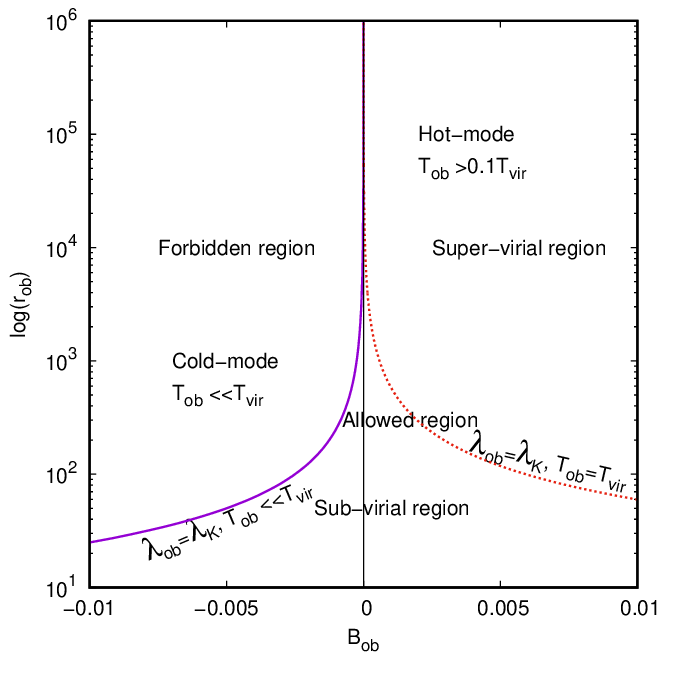}
\caption{OBC-plane: Represents variation of $\bob$ with $\rob$. The region left side of the black thin line ($\bob=0$) represents cold-mode inflow gases and the solid-violet line represents $\amob=\amk$ with $\thob<<\thvir$. 
The region's right side of $\bob=0$ represents hot-mode inflow gases and the dotted-red line represents $\amob=\amk$ with $\thob=\thvir$. An upper left region of the solid-violet line is named a forbidden region because 
these energies can not be possible for inflowing gases. The rest of the region, the right side of the solid-violet line is named an allowed region which can be further divided into sub- and super-virial regions.
}
\label{fig:1}       
\end{figure}
This study has focused on the investigations of the nature of accretion flows by changing the initial temperature of the inflowing gases at the outermost boundary locations (OBLs) around the black hole. Recently, we have defined outer boundary conditions (OBCs) for the accretion flows which can represent the external gas environment around the accreting object \cite{ky21}. Those OBCs can be presented on the plane of local specific energy ($\bob$) of the inflowing gases and OBLs ($\rob$) 
and we named this plane as OBC-plane. The details of the OBC-plane are presented in Fig. \ref{fig:1}. The $\bob$ of the gas is defined by using the canonical Bernoulli parameter in Eq. \ref{be.eq}. 
The definition of $\bob$ at $\rob$ is that the initial inflow velocity $v\sim0$, initial temperature $\thob$ varies $\thvir>>\thob\sim\thvir$, and initial specific AM $\amob=\amk$ of the gases. Here, $\amk$, and $\thvir$ are Keplerian AM, and dimensionless Virial temperature at $\rob$, respectively. {The virial temperature is used to show the variation of the $\bob$ in the OBC-plane. However, the initial temperature of inflowing gas at the OBL is also a free parameter.}
The Virial temperature of a gas at OBL is defined by following the Virial Theorem, and energy equipartition theorem for gases, so $\thvir$ is expressed as 
\begin{equation}
\frac{3}{2\mu}kT_{\rm vir} = \frac{1}{2}\frac{GMm_H}{R-\rs}\Rightarrow
\thvir=\frac{\mu}{6}\frac{(1+m_p/m_e)}{r-1}\, ,
\label{eq:vir}
\end{equation}
here, $1.5k\tvir$ is the average kinetic energy of the particle, $R=r\rs$, and $\thvir=k\tvir/(m_{e^-}c^2)$. $k$, $\mu$, and $T_{\rm vir}$ are the Boltzmann constant, mean molecular weight of the gas, and Virial temperature in the physical unit.
The Virial sound speed can be defined using Eq. \ref{as.eq} as 
\begin{equation}
a_{vir}=\sqrt{\frac{2\gamma\thvir}{\tilde{t}}}=\sqrt{\frac{\mu\gamma}{3(r-1)}},
\label{avir.eq}
\end{equation}
here we used $\mu=1$, $\gamma=5/3$ and $r=\rob$. So the $\bob$ versus $\rob$ is plotted in Fig. \ref{fig:1}. 

Primarily, the whole OBC-plane can be divided into forbidden and allowed energy regions of the inflowing gases at the OBLs as shown in Fig. \ref{fig:1}. A solid violet line in the cold-mode region is drawn with conditions $\amob=\amk$ and $\thob<<\thvir$ at $\rob$ and it separates the forbidden and allowed energy regions in the OBC-plane. A region on the left side of this line is the forbidden energy region, and the right side region is the allowed energy region for the inflowing gases. In the forbidden region, the energies are not possible for the inflowing gases.  The allowed energy region can be further divided into cold-mode and hot-mode inflowing gases/regions, and both regions are separated by the $\bob=0$ solid black thin line. The hot-mode region is again divided into sub- and super-virial regions, and it is separated by the dotted red line which is generated by using conditions $\amob=\amk$ and $\thob=\thvir$. {This dotted red curve not only separates virial and super virial regions is also provides the qualitatively the same variation of the $\bob$ corresponding to the types of accretion solutions for a particular viscosity parameter. This we will show in Fig \ref{fig:4} for the shocked flows only. Moreover, the OBCs corresponding to the three CP solutions are mostly found in the super virial region and this kind of detailed study will be reported elsewhere in the future.}
 The cold-mode region has only sub-virial temperature.

\section{Results and Discussions}
\label{sec:4}
We scanned the whole OBC-plane and found five types of global accretion solutions that are represented by some of their properties in the following Figs (\ref{fig:2} - \ref{fig:5}).
Interestingly, the cold-mode gas with $\bob<0$ gives only global advection-dominated accretion flow (ADAF) \cite{ny94,nkh97}. However, the hot-mode gas with $\bob>0$ gives other four types of global accretion solutions with different initial temperatures or $\bob$ at the $\rob$.
A Fig. \ref{fig:2} represents  $\lambda$ distributions for the five global accretion solutions with different initial temperatures or $\bob$ of inflowing gases at the same OBL as mentioned in the caption. 
A solid violet line represents  ADAF. A dashed green line represents ADAF-thick flow \cite{lgy99}. Both solutions have a single inner CP, but the ADAF-thick flow is geometrically thicker than the ADAF.  Dotted, dash-dotted, and dash-double dotted curves represent W-type (3-CP but passes only through inner CP), shocked (3-CP), and Bondi-types (single outer CP) accretion solutions, respectively \cite{c89,c96}.  Interestingly, the ADAF-thick, W-type, shocked and Bondi-type accretion solutions are only generated by the hot-mode gas $\bob>0$ at OBL. Moreover, it is seen that the AM transportation due to viscosity is more efficient with higher temperatures of inflowing gases at the OBL. Thus, the ADAF solution with the lowest temperature (cold-mode gas) at OBL has the highest AM distribution, and the Bondi-type with the highest temperature (hot-mode gas) at OBL has the lowest AM distribution. Here, it can be noted that the low AM distribution allows the formation of outer CP in the solutions, such as in the shocked and Bondi-type flows. So, the initial temperature of inflowing gases should be high enough because $W_{r\phi}\propto a$ or $\Theta$ so that the outward AM transportation by the viscosity can become efficient to reduce the AM distribution in the disk to make the outer CP if the inflowing gases at $\rob$ has Keplerian AM. 
\begin{figure}[t]
\sidecaption[t]
\includegraphics[angle=0,scale=.66]{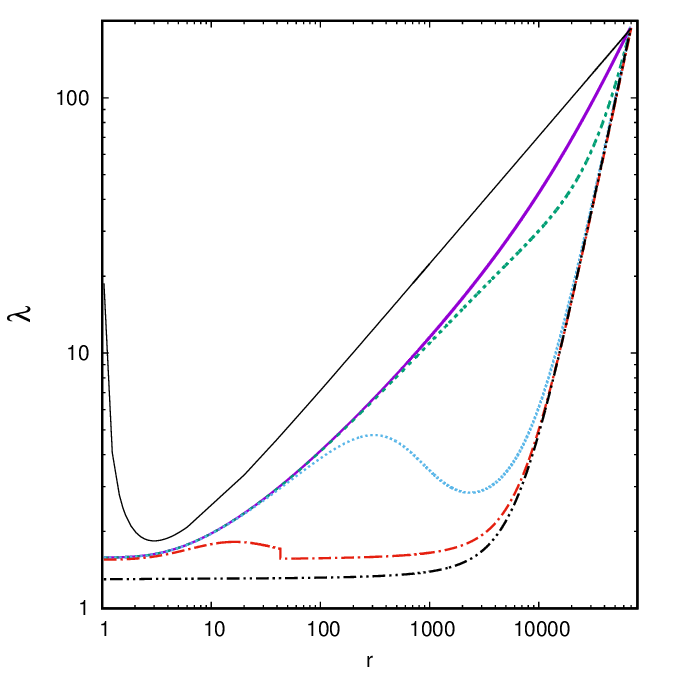}
%
%
\caption{Variations $\lambda$ versus $r$ for five types of accretion solutions with same $\rob\sim6.92e4, ~\amob=\amk$ and $\alpha=0.05$ but different initial temperature or $\bob=-2.70e-6$ (solid-violet line, ADAF), $1.23e-5$ (dashed-green, ADAF-thick), $1.07e-4$ (dotted-blue, W-type), $2.43e-4$ (dashed-dotted red, Shock) and $2.65e-4$ (dashed-double dotted black, Bondi-type) at the OBL. A solid black curve represents Keplerian AM distribution.}
\label{fig:2}       
\end{figure}

A Fig. \ref{fig:3} shows the variation of the sound speed ($a$) with radial distance ($r$) corresponding to the same solutions of Fig. \ref{fig:2} with the same line styles. 
A solid black curve represents the variation of the viral sound speed ($a_{vir}$), which is expressed in Eq. \ref{avir.eq}. The cases of Fig. \ref{fig:3} show that the $a$ of W-type, shocked, and Bondi-types solutions at OBL exceed the $a_{vir}$, however the ADAF-thick solution has sub-virial sound speed in hot-mode inflow gases.
 Usually, we consider the Virial temperature or corresponding sound speed as the maximum temperature/sound speed of the gases in the gravitationally bound system. However, the super-virial temperature of the gases can be achieved through external heating, such as radiative heating. The sub- and super-virial temperature positions can be changed for particular solutions with changing flow parameters, such as $E, \lambda_0$, and $\alpha$. Here, particularly we are intereseted in shocked solutions because of two reasons, 1. The parameter space of shocked flow is small compared to other solutions, and it shrinks with increasing $\alpha$ \cite{kc13,ky21}, and 2. Shocked solutions can have many importance, such as high energy nonthermal radiations, outflow, and QPOs generation \cite{ct95,mc10,lck16}. So, the OBC of the shocked flow is explored with different viscosity parameters and presented in Fig. \ref{fig:4}.  In this figure, we plotted the initial temperature ($\tob$ in the unit of $K$) of the inflowing gases at OBL in the left panel. Here, $\tob$ is the proton temperature of the gas. We plotted it because electron and proton temperatures are almost the same around the OBL in the two-temperature accretion flows \cite{scl20}. 
 Here, we plotted three curves for $\alpha=0.1,~0.2, ~0.25$ and shocked parameter space for higher $\alpha>0.25$ is almost vanished.
 This figure shows that the $\tob$ for the shocked flow is much higher than the $\tvir$. So this can raise questions about the existence of the steady-state shocks in accretion flows with $\amob=\amk$ at $\rob$ if the super-virial temperature of the inflowing gases at OBL is not found around the BHs. { For example, the shocked flow at $\rob\sim10^5\rs$ in the left panel of Fig \ref{fig:4} has temperature $\tob>10^8$K and this temperature at such radius is not found yet in observational studies in the BH accretion system.}
\begin{figure}[t]
\sidecaption[t]
\includegraphics[angle=0,scale=.65]{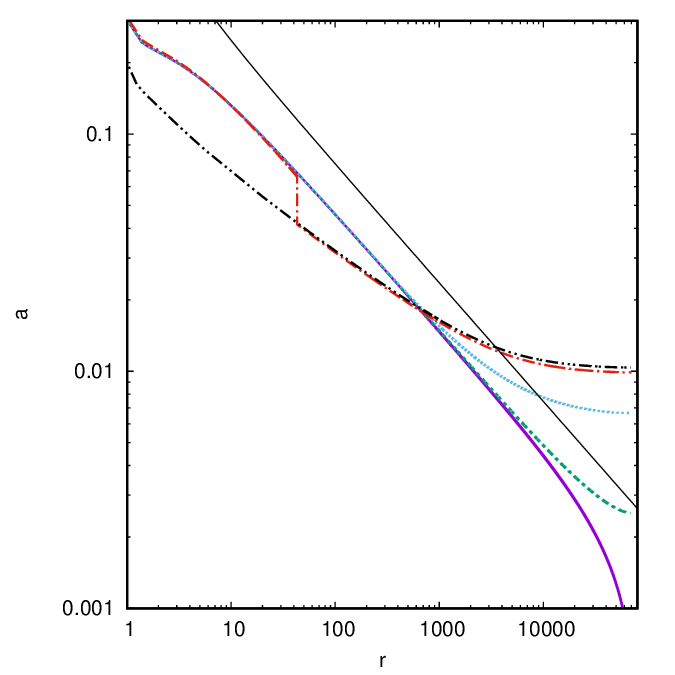}
%
%
\caption{Variations of $a$ versus $r$ with same five solutions as in Fig. \ref{fig:2}. A solid black curve represents the variation of the Virial sound speed ($a_{vir}$).}
\label{fig:3}       
\end{figure}

\begin{figure}[t]
\includegraphics[angle=0,scale=.5]{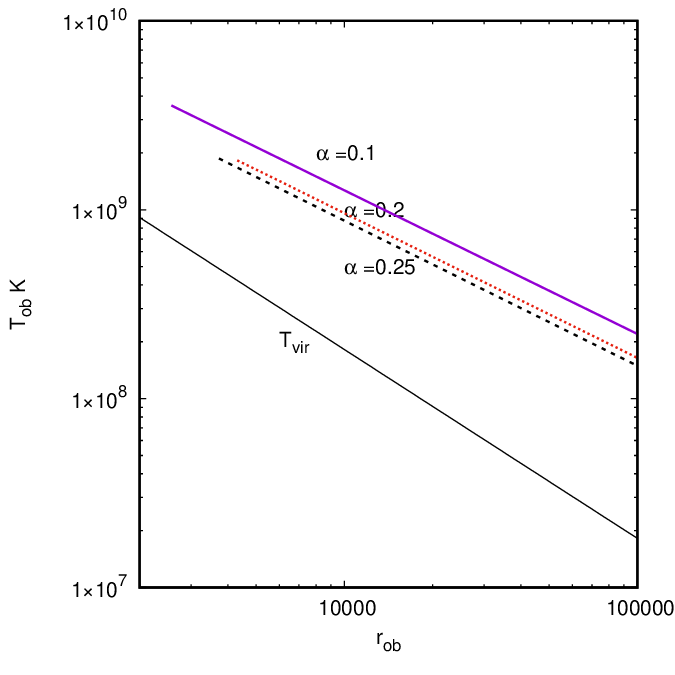}
\includegraphics[angle=0,scale=.5]{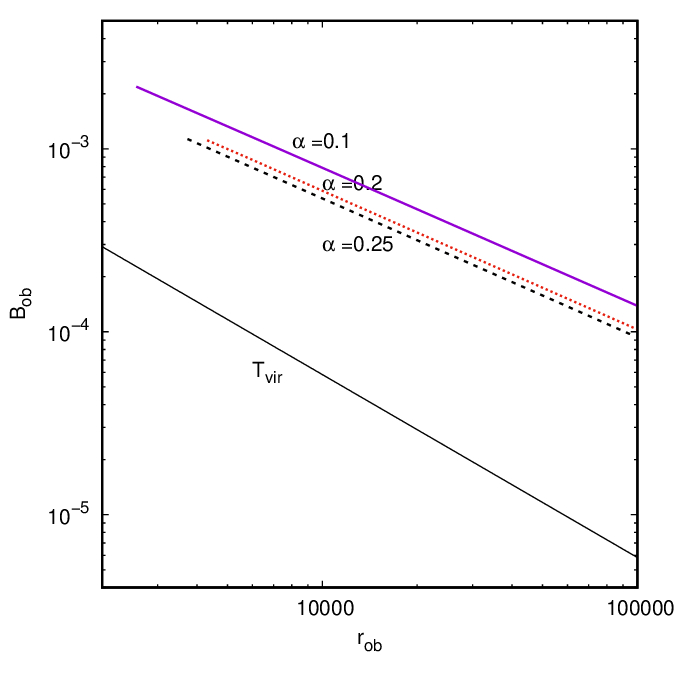}
%
%
\caption{Variation of $\tob$ (in physical unit) with $\rob$ for global shocked accretion solutions for different $\alpha=0.1$ (solid-violet line), $0.2$ (dotted-red line), and $0.25$ (dashed-black line). A solid thin black line represents the variation of $\tvir$. The $\rob$ is calculated when $\amob=\amk$. In the right panel, we plotted $\bob$ corresponding to the same solutions of the left panel. The right panel represents a part of the hot-mode region of the OBC-plane. 
}
\label{fig:4}       
\end{figure}
{Interestingly, in Fig. \ref{fig:4}, the variations of the $\tob$ for shocked flow follow almost the same trends as the variation of $\tvir$ with distance in the gravitational system. However, the slope of the decreasing $\tob$ of the shocked flows with $\rob$ is flatter than the variation of the $\tvir$ in especially large radii. Now we checked the variation of the $\bob$ corresponding to the shocked solutions in the OBC-plane. So we have plotted $\rob$ versus $\bob$ in the right panel of Fig. \ref{fig:4} with $\bob$ corresponding to the $\tvir$. This panel of Fig. \ref{fig:4} represents only a part of the hot-mode region of the OBC plane that focuses on the OBCs corresponding to the shocked flows. We again found that the variation of $\bob$ for shocked flow is almost the same as the variation of the $\bob$ corresponding to the $\tvir$. Thus, the dotted red curve in Fig \ref{fig:1} gives a qualitative understanding of the variation of the $\bob$ corresponding to the OBCs of the global accretion solutions for a particular viscosity parameter with $\amob=\amk$ in the OBC plane. Here, we have plotted $\bob$ for only the shocked flows. Interestingly the variation of the $\bob$ with $\rob$ is also the same for the other kinds of accretion solutions, however, the location of those curves is different in the OBC plane.
}

\begin{figure}[t]
\sidecaption[t]
\includegraphics[angle=0,scale=.65]{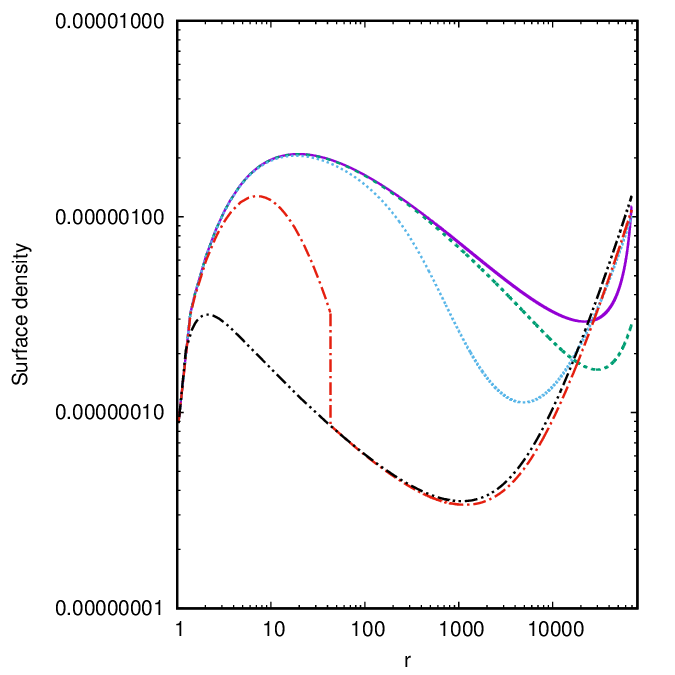}
%
%
\caption{Variations of the surface density ($\Sigma$) with radial distance ($r$) for five different accretion solutions of Fig. \ref{fig:2}. }
\label{fig:5}   
\end{figure}
Fig. \ref{fig:5} represents the variation of the surface density ($\Sigma=2\rho h$) for the same five accretion solutions. The variation of the surface density can represent the optical depth of the disk with a multiplication of Thompson scattering cross-section.  The change in optical depth can affect the radiative emission of the disk. This figure shows that the surface density of the disk can change more than one order of magnitude by only changing the initial temperature of the inflowing gases at OBL. So we can use the $\tob$ at the OBL as a parameter as the mass accretion rate is used in the spectral modeling studies. 
\section{Conclusions}
\label{sec:5}
In this study, we have investigated the OBC-plane (Fig. \ref{fig:1}) which has shown two types of the behavior of variations of the $\rob$ with $\bob$ and gives cold-mode and hot-mode inflowing gases at the OBLs. Both types of gases have given us different types of accretion solutions. Interestingly, $\bob<0$ gives only global ADAF solutions \cite{nkh97}. $\bob>0$ gives global ADAF-thick \cite{lgy99}, W-types, shocked and Bondi-type solutions \cite{c89}.  {This study also explains why some researchers find accretion solutions with multiple sonic points \cite{c89,kc13} while others find single sonic point ADAF solutions \cite{ny94,nkh97}.}
We have investigated these solutions by keeping $\rob$ and $\alpha$ fixed but only changing the initial temperature of the inflowing gases at  $\rob$ as shown in Fig. \ref{fig:2}. Here, we have found that the ADAF has the highest AM distribution and Bondi-type has the lowest AM distribution. Similarly, these solutions have different distributions of the surface density (Fig.\ref{fig:5}) which can affect the radiative emissivities of the disk. So we proposed that the initial temperature of the inflowing gases at OBLs can be a parameter for the spectral modeling studies of the accreting sources. In this study, we have also found that the global shocked and Bondi-type solutions have been generated from the super-virial temperature of the inflowing gases at the OBLs. So this result can be a demerit for the having of the steady state shocks in the BH accretion systems unless they have super-virial hot gas in their outer environment. However, it can be possible that the super-virial hot gas can be generated in the outer environment by radiative heating. 

\begin{acknowledgement}
I would like to express my gratitude to the organizers of the International Symposium on Recent Developments in Relativistic Astrophysics at SRM University, Gangtok, India.
\end{acknowledgement}
\ethics{Competing Interests}{Author has no competing interest.
}



%
%
%


\end{document}